\documentclass[10pt,aps,pra,twocolumn,showpacs]{revtex4-1}
\usepackage{xr}

\usepackage{amsmath}
\usepackage{amsfonts, amssymb,amsxtra}
\usepackage[]{graphicx}
\usepackage{hyperref}
\usepackage[figure,table]{hypcap}               
\hypersetup{
pdftitle = {},
pdfsubject = {},
pdfauthor = {},
pdfkeywords = {},
pdfcreator = {},
pdfproducer = {LaTeX with hyperref},
colorlinks = true,
linkcolor = red, 
anchorcolor = red,
citecolor = red, 
filecolor = red, 
menucolor = red, 
pagecolor = red, 
urlcolor  = red, 
breaklinks = true,
pdfstartview = FitV,
pdfhighlight = /I,
pdfpagelayout = OneColumn,
hypertexnames=true 
}

\newcommand{\ket}[1]{|~\!#1~\!\rangle}

\newcommand{\braopket}[3]{\langle#1|#2|#3\rangle}
\newcommand{\av}[1]{\langle #1 \rangle}

\begin{document}
\title{Universality in dissipative Landau-Zener transitions}

\author{Peter P. Orth,$^{1}$ Adilet Imambekov,$^{2}$ and Karyn Le Hur$^{1}$}
\affiliation{$^{1}$Department of Physics, Yale University, New Haven, Connecticut 06520, USA \\
$^{2}$Department of Physics and Astronomy, Rice University, Houston, Texas, 77005, USA}
\date{\today}
\begin{abstract}
We introduce a random variable approach to investigate the dynamics of a dissipative two-state system. Based on an exact functional integral description, our method reformulates the problem as that of the time evolution of a quantum state vector subject to a Hamiltonian containing random noise fields. This numerically exact, non-perturbative formalism is particularly well suited in the context of time-dependent Hamiltonians, both at zero and finite temperature. As an important example, we consider the renowned Landau-Zener problem in the presence of an Ohmic environment with a large cutoff frequency at finite temperature. We investigate the 'scaling' limit of the problem at intermediate times, where the decay of the upper spin state population is universal. Such a dissipative situation may be implemented using a cold-atom bosonic setup.
\end{abstract}
\maketitle
\section{Introduction}
\label{sec:introduction}
A two-level system is never completely isolated resulting in dissipation, decoherence and entanglement~\cite{lehur_entanglement_spinboson}.
Therefore, one primary task for experimentalists is to manipulate and read out the internal state of the dissipative two-level system (qubit) with a high fidelity. Often, this can be achieved by sweeping the two energy levels through an avoided crossing, a situation that occurs in a variety of physical areas such as molecular collisions~\cite{Child_molecular_collision_theory_book}, chemical reaction dynamics~\cite{Nitzan_chem_dyn_in_condensed_phase_book}, molecular nanomagnets~\cite{W.Wernsdorfer04021999}, quantum information and metrology~\cite{Wallraff_Nature_2004, sillanpPRL, OrlandoLevitov_AmplitSpectr_Nature_2008,Manucharyan10022009}.
For a constant crossing speed $v$ this is known as the Landau-Zener problem~\cite{landau_lz,zener_lz,stueckelberg_lz,majorana_lz} which can be solved exactly in the absence of dissipation. Naturally, it is important to know the effect of the dissipative universe on the probability $p(t)$ for the spin to remain in its initial state at time $t$~\cite{PhysRevLett.62.3004,PhysRevB.43.5397,PhysRevB.57.13099,nalbach:220401}. Exact results~\cite{wubs:200404,saito:214308} are only available at zero temperature and in the limit $t\rightarrow +\infty$, where the energy difference $\epsilon$ of the two spin states is much larger than the bandwidth $\omega_c$ of the environmental bath. Typically however, $\omega_c$ is much larger than the tunneling coupling between the two states $\Delta$. Here, we rather focus on the experimentally relevant ``scaling'' regime at intermediate times, where the spin energies have not completely traversed the bath's energy band: $\Delta < \epsilon = v t < \omega_c $ with $v>0$. To resolve the dissipative spin dynamics, we develop a powerful numerically exact stochastic Schr\"odinger equation formalism (SSE). Compared to earlier SSE approaches~\cite{Kleinert1995224,PhysRevLett.80.2657,PhysRevLett.88.170407,PhysRevLett.82.1801}, our method allows easier exact consideration of initial spin-bath correlations, which are crucial in the Landau-Zener context. It may also be applied to other many-body environments that can be represented in the form of a Coulomb gas such as the Kondo model.~\cite{imambekov:063606,0022-3719-4-5-011} 

We prove that $p(t)$ exhibits a universal decay in the intermediate (scaling) regime due to phonon assisted spin transitions. The size of the jump at the level crossing decreases for increasing dissipation and $p(t)$ converges to the infinite time value only when $t \sim \omega_c/v$. 
We also derive an approximate analytical decay formula valid for slow sweeps at zero temperature, which agrees well with our numerical results. 


\section{Model and Notations} 
\label{sec:model-notations}
Specifically, we study a two-level system coupled to a bath of harmonic oscillators (the spin-boson Hamiltonian)~\cite{RevModPhys.59.1,weissdissipation}
\begin{equation}
  \label{eq:1}
  \frac{H}{\hbar} = \frac{\Delta}{2} \sigma^x + \frac{\epsilon}{2} \sigma^z + \frac{\sigma^z}{2} \sum_k \lambda_k (b_k^\dag + b_k) + \sum_k \omega_k b^\dag_k b_k\,.
\end{equation}
Here, $\sigma^{x,z}$ are the Pauli matrices, $\Delta$ is the bare tunneling coupling and $\epsilon$ the detuning. The bosonic oscillator operators have frequencies $\omega_k$ and coupling constants $\lambda_k$. We express the components of the reduced spin density matrix $\rho(t)$ using functional integrals~\cite{RevModPhys.59.1,weissdissipation}
\begin{equation}
  \label{eq:2}
  \rho(\sigma_f,\sigma'_f;t) = \int \mathcal{D} \sigma(\cdot) \int \mathcal{D} \sigma'(\cdot) \mathcal{A}[\sigma] \mathcal{A}^*[\sigma'] F[\sigma,\sigma']\,,
\end{equation}
where $\mathcal{A}[\sigma]$ is the amplitude for the spin to follow the path $\sigma(t)$ in the absence of the bath, and $F[\sigma, \sigma']$ is the real-time influence functional of the bath
\begin{align}
  F[\sigma,\sigma'] &= \exp \Bigl[ - \frac{1}{\pi \hbar} \int_{t_0}^t ds \int_{t_0}^{s} ds'  \{- i L_1(s-s') \xi(s) \eta(s') \notag \\
  & \quad + L_2(s-s') \xi(s) \xi(s') \} \Bigr]\,, \label{eq:3}
\end{align}
written in terms of symmetric and antisymmetric spin paths $\eta(s) = \frac12 [ \sigma(s) + \sigma'(s)]$ and $\xi(s) = \frac12 [ \sigma(s) - \sigma'(s)]$, respectively. The kernel functions $L_1(t) = \int_0^\infty d\omega J(\omega) \sin \omega t$ and $L_2(t) = \int_0^\infty d\omega J(\omega) \cos \omega t \coth \hbar \omega/2 k_B T$ are determined by the bath spectral function $J(\omega) = \hbar \pi \sum_k \lambda_k^2 \delta(\omega - \omega_k)$ and the temperature $T$.

At time $t_0\rightarrow - \infty$, the spin-bath interaction is first turned on, but    
the spin is held fixed in position $\sigma_i$ for $t_0<t\leq0$. The spin paths $\{\sigma(t), \sigma'(t)\}$ in Eq.~\eqref{eq:2} are constrained to $\sigma(t)=\sigma'(t) = \sigma_i$ for $t \leq 0$ and to $\sigma(t_f)=\sigma_f, \sigma'(t_f)=\sigma'_f$. At $t=0$, the bath is in the shifted canonical equilibrium state.
For positive times, the spin jumps between the states $\{\ket{\!\!\uparrow}, \ket{\!\!\downarrow}\}$ and the spin double path occurring in Eq.~\eqref{eq:2} can thus be regarded as a single path between the four states $\{\ket{\!\!\uparrow \uparrow}, \ket{\!\!\uparrow \downarrow}, \ket{\!\!\downarrow \uparrow}, \ket{\!\!\downarrow \downarrow}\}$. If the path starts and ends in a diagonal (``sojourn'') state $\{\ket{\!\!\uparrow \uparrow}, \ket{\!\!\downarrow \downarrow}\}$ and  makes $2n$ transitions at times $t_1<t_2<\ldots< t_{2n}$ along the way, it can be parametrized as $\xi(t) = \sum_{j=1}^{2n} \Xi_j \Theta(t - t_j)$ and $\eta(t) = \sum_{j=0}^{2n} \Upsilon_j \Theta(t - t_j)$. The variables $\{\Xi_1, \ldots, \Xi_{2n}\} = \{\xi_1, - \xi_1, \dots, - \xi_n\}$ embody the $n$ off-diagonal (``blip'') parts of the path between the times $t_{2m-1}$ and  $t_{2m}$ ($m=1,\dots,n$), and characterize the time spent by the path in the states $\{\ket{\!\!\uparrow \downarrow}, \ket{\!\!\downarrow \uparrow}\}$ such that $\xi(t) = \pm 1, \eta(t) = 0$. The variables $\{\Upsilon_0, \ldots, \Upsilon_{2n}\} = \{\eta_0, -\eta_0, \ldots, \eta_n\}$ describe the $n+1$ diagonal (sojourn) parts in the time period $(t_{2m},t_{2m+1})$ during which $\eta(t) = \pm 1, \xi(t) = 0$ (here, we have $m = 0, \ldots, n$ and $t_{2n+1} \equiv t_f$). The path's boundary conditions then specify $\eta_0$ and $\eta_n$.

Inserting this general spin path $\xi(t), \eta(t)$ into Eq.~\eqref{eq:3} and performing the time integrations yields $F_n[\Xi_j, \Upsilon_j, t_j] = \mathcal{Q}_1 \mathcal{Q}_2$ with
\begin{align}
  \label{eq:4}
  \mathcal{Q}_1 &= \exp \Bigl[ \frac{i}{\pi \hbar} \sum_{j>k\geq 0}^{2n} \Xi_j \Upsilon_k Q_1(t_j - t_k) \Bigr]\,, \\
\label{eq:5}
  \mathcal{Q}_2 &= \exp \Bigl[ \frac{1}{\pi \hbar} \sum_{j>k\geq1}^{2n} \Xi_j \Xi_k Q_2(t_j - t_k)  \Bigr]\,,
\end{align}
where $Q_{1,2}$ are the second integrals of $L_{1,2}$. 
The free spin-path amplitudes $\mathcal{A}[\sigma]\mathcal{A}^*[\sigma']$ give a factor $i\xi\eta \Delta/2$ to switch from a sojourn state $\eta$ to a blip state $\xi$ (and vice versa) as well as a bias-dependent phase factor $H_n=\exp[ i \sum_{j=1}^{2n} \Xi_j s(t_j)]$ with $s(t) = \int_0^{t} dt' \epsilon(t')$. Altogether, the probability $p(t)=\rho(\ket{\!\!\uparrow}, \ket{\!\!\uparrow};t)$ to find the system in state $\ket{\!\!\uparrow}$ at time $t$ takes the form,
\begin{equation}
\begin{split}
    \label{eq:6}
    p(t) = 1 + \sum_{n=1}^{\infty} \Bigl(\frac{i \Delta}{2}\Bigr)^{2n} \int_0^t dt_{2n} \cdots \int_0^{t_2} dt_1 \sum_{\{\xi_j, \eta_j\}} F_n H_n.
\end{split}
\end{equation}

\section{Random Variables}
\label{sec:random-variables}
We now proceed and decouple the terms bilinear in the blip and sojourn variables by Hubbard-Stratonovich transformations. Such a decoupling is useful since Eq.~\eqref{eq:6} has the Coulomb gas structure.~\cite{imambekov:063606}. Our formalism may thus be applied to other models which allow a Coulomb gas representation such as the Kondo model~\cite{0022-3719-4-5-011}.
The resulting expression then suggests that $p(t)$ can be obtained as a statistical average of a stochastic Schr\"odinger equation~\cite{kubo:174, Kleinert1995224, PhysRevLett.80.2657,PhysRevLett.88.170407,PhysRevLett.82.1801,imambekov:063606}.

For definiteness, we will now focus on the case of an Ohmic bath with spectral function $J(\omega) = \eta \omega \exp( - \omega/\omega_c)$. It contains the viscosity coefficient $\eta$ and a high-frequency cutoff $\omega_c$, and we also introduce the dimensionless dissipation parameter $\alpha=\eta/2 \pi \hbar$. We like to emphasize that our method is able to solve for the system's dynamics at any temperature $T$. 
The bath correlation functions read $Q_1(t) = \eta \tan^{-1}(\omega_c t)$ and $Q_2(t) = \frac{\eta}{2} \ln(1 + \omega_c^2 t^2) + \eta \ln \Bigl[ \frac{\hbar}{\pi k_B T t} \sinh \frac{\pi k_B T t}{\hbar} \Bigr]$~\cite{RevModPhys.59.1,weissdissipation}. 

In fact, to apply a Hubbard-Stratonovich transformation to Eq.~\eqref{eq:5}, we need to write $Q_2(t_j-t_k)$ in a factorized form $Q_2(t_j-t_k) = \frac{\eta}{2} [G_0 + \sum_{m=1}^{m_{\max}} G_m \Psi_m(t_j) \Psi_m(t_k)]$. Since the kernel is translationally invariant, this can be achieved by a Fourier series expansion. To obtain only negative Fourier coefficients, we rather expand $\Tilde{Q}_2(\tau) = Q_2(\tau) - Q_2(2) = \frac{\eta}{2} [g_0 + \sum_{m=1}^{m_{\max}/2} g_m \cos \frac{m \pi \tau}{2}]$, where we introduced the rescaled time $\tau = t/t_{\text{max}}$, with $t_{\max}$ being the final time of our numerical simulation. Thus, the coefficients are $G_0 = g_0 + \frac{2}{\eta}Q_2(2)$, $G_{2k-1}$$=$$G_{2k}$$=$$g_k$$<0$, and the trigonometric functions read $\Psi_{2k-1}$$=$$\cos \frac{k \pi \tau}{2}$, $\Psi_{2k}$$ =$$ \sin \frac{k \pi \tau}{2}$, where $k$$=$$1,\ldots, m_{\max}/2$.
Decoupling the blip variables by $m_{\max}$ Hubbard-Stratonovich transformations then results in
\begin{equation}
  \label{eq:7}
\mathcal{Q}_2 = e^{- n \alpha [\frac{2}{\eta} Q_2(2) + G]}\int d\mathcal{S} \exp\Bigl[ i \sum_{j=1}^{2n} \Xi_j h(\tau_j) \Bigr]\,,
\end{equation}
where the sum $G$$=$$\sum_{m=0}^{m_{\max}/2} g_m$ is equal to $[-\frac{2}{\eta}Q_2(2) ]$ for $m_{\max}$$ \rightarrow$$\infty$, the integration over the Gaussian distributed Hubbard-Stratonovich variables reads $\int d\mathcal{S} = \prod_{m=1}^{m_{\max}} \int_{-\infty}^{\infty} \frac{ds_m}{\sqrt{2\pi}} e^{- s_m^2/2}$, and we have introduced the real function $h(\tau) = \sum_{m=1}^{m_{\max}} s_m \sqrt{ - \alpha G_m} \Psi_m(\tau)$.

We can proceed similarly with $Q_1$ after separating it into a symmetric $Q_1(|t|)$ and an antisymmetric part $Q_1(t)$ in order to extend the sum to $j\leq k$. On the other hand, for zero detuning $\epsilon=0$ and $\alpha<1/2$~\cite{RevModPhys.59.1,weissdissipation}, one can safely approximate $Q_1(t) \approx \eta \pi/2$. This approximation becomes exact for $\Delta/\omega_c \rightarrow 0$ since the main contribution to the functional integral of Eq.~\eqref{eq:6} stems from spin flips with time separations larger than $\omega_c^{-1}$. The finite bias case $\epsilon \neq 0$ requires more consideration of the first sojourn as it accounts for the spin-bath preparation, which affects the long-time behavior of $p(t)$~\cite{saito:214308} (see below).
\begin{figure*}[t!]
  \centering
 \includegraphics[width=.7\linewidth]{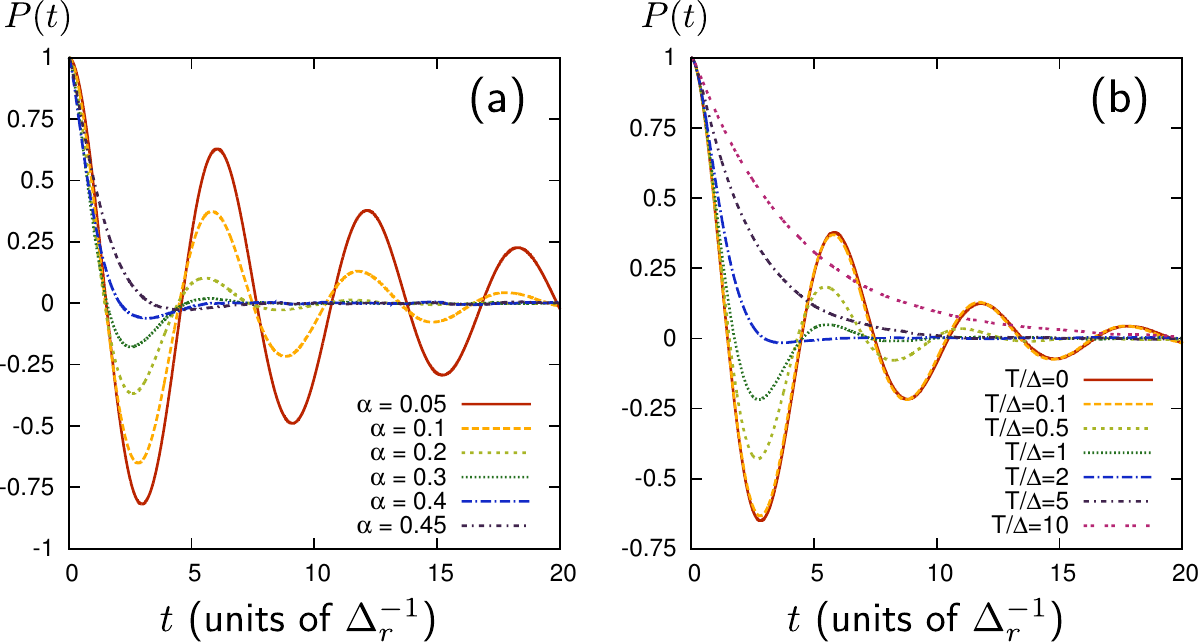}
  \caption{(Color online): (a) $P(t)$ as a function of $t$ for various values of $\alpha$, $\Delta=1$, $\omega_c=100$, $\epsilon=0$ and $T=0$. We checked that for a given $\alpha$ curves corresponding to different $\omega_c/\Delta \gg 1$ scale on top of each other in units of the renormalized tunneling rate $\Delta_r = \Delta (\frac{\Delta}{\omega_c})^{\alpha/(1-\alpha)}$. Quality factor $\Omega/\gamma$ of damped oscillations agrees with prediction $\Omega/\gamma = \cot \frac{\pi \alpha}{2 (1 - \alpha)}$ from Refs.~\cite{RevModPhys.59.1,weissdissipation,PhysRevLett.80.4370}. Results are obtained with $m_{\max} = 3000$, $N=5\cdot 10^4$. 
(b) $P(t)$ for different temperatures $T$ (here, $\hbar = k_B =1$), dissipation strength $\alpha = 0.1$, and other parameters as in (a).}
  \label{fig:1}
\end{figure*}
For $\epsilon = 0$, Eq.~\eqref{eq:6} reads
\begin{equation}
  \label{eq:8}
\begin{split}
  &p(\tau) = 1 + \int d\mathcal{S} \sum_{n=1}^{\infty} \Bigl( \frac{ i \Delta t_{\max} e^{- \frac{\alpha}{2} [\frac{2}{\eta} Q_2(2) + G]}}{2} \Bigr)^{2n} \int_0^\tau d\tau_{2n}\\ &\times \cdots \int_0^{\tau_2} d\tau_1 \sum_{\{\xi_j, \eta_j\}} \exp [ i \pi \alpha \sum_{k=0}^{n-1} \eta_k \xi_{k+1}] \prod_{j=1}^{2n} \exp[i \Xi_j h(\tau_j)]\,.
\end{split}
\end{equation}
Without the summation over blip and sojourn variables $\{\xi_j, \eta_j\}$, this expression has the form of a time-ordered exponential, averaged over the random variables $\{s_m\}$. This summation, however, can be incorporated into a product of matrices in the vector space of states $\{\ket{\!\!\uparrow \uparrow}, \ket{\!\!\uparrow \downarrow}, \ket{\!\!\downarrow \uparrow}, \ket{\!\!\downarrow \downarrow}\}$, which have the form~\cite{imambek_jetp_02}
\begin{equation}
  \label{eq:9}
  V \!=  \!A \!\begin{pmatrix} 0 & e^{- i h(\tau)} & - e^{i h(\tau)}& 0  \\ e^{i \pi \alpha}e^{i h(\tau)} & 0 & 0 & - e^{- i \pi \alpha} e^{ i h(\tau)} \\ - e^{- i \pi \alpha}e^{ - i h(\tau)}  & 0 & 0 & e^{i \pi \alpha}e^{- i h(\tau)} \\ 0 & - e^{- i h(\tau)} & e^{i h(\tau)} & 0 \end{pmatrix},
\end{equation}
with $A = \frac12 (\Delta t_{\max} e^{- (\alpha/2) [ (2/\eta) Q_2(2) + G]})$.
Then, Eq.~(\ref{eq:8}) becomes $p(\tau) = \int d \mathcal{S}  \braopket{\Phi_f}{T e^{-i \int_0^\tau ds V(s) }}{\Phi_i}$
which can be calculated by solving the stochastic Schr\"odinger equation
\begin{equation}
  \label{eq:10}
  i \frac{\partial }{\partial \tau} \ket{\Phi(\tau)} = V(\tau) \ket{\Phi(\tau)},
\end{equation}
with initial and final conditions $\ket{\Phi_{i,f}}=(1,0,0,0)^T$ for $N$ different realizations of the noise variables $\{s_m\}$. Averaging the results gives $p(\tau) = \frac{1}{N} \sum_{k=1}^N \Phi_1^{(k)}(\tau)$, where $\Phi_1(\tau)$ is the first component of $\ket{\Phi(\tau)}$. 
Other components of the density matrix (\ref{eq:2}) can be obtained using different initial and final conditions. 
In fact, the differential equations obey the additional symmetries $\text{Im} \Phi_1=0$, $\Phi_3^*= \Phi_2$ and $\Phi_4 = 1-\Phi_1$, such that only three real-variables are independent. Since the evolution is unitary (for $\epsilon=0$) and $\Phi_1^2 + 2 |\Phi_2|^2 = 1$ is an integral of motion, we can introduce a classical unit-length spin ${\bf S} = (\sqrt{2} \text{Re} \Phi_2, \sqrt{2} \text{Im} \Phi_2, \Phi_1)$ that evolves according to $d{\bf S}/d\tau = {\bf H} \times {\bf S}$ in a random magnetic field ${\bf H} = (\cos h(\tau), \sin h(\tau), 0)$, and from which we find $p(\tau) = \frac{1}{2} (1 + \langle S^z(\tau) \rangle)$.
Hence, the time-evolution of a dissipative quantum spin can be formulated as that of a classical spin in a random magnetic field. The quantum nature of the problem is hidden in the fact that spin rotations about different axes do not commute and through the averaging over random field configurations. 

\section{Applications}
\label{sec:applications}
\subsection{Spin dynamics at zero detuning}
\label{sec:spin-dynamics-at}

To prove the feasibility of our method, we have computed the spin dynamics for zero detuning in the range $0 < \alpha < 1/2$ for different temperatures $T$. We express $T$ in units of $\Delta$ (hereafter we set $\hbar = k_B = 1$). Results for $P(t) = 2 p(t) - 1$ in Fig.~\ref{fig:1} exhibit damped oscillations with the correct renormalized tunneling frequency of order $\Delta_r = \Delta (\frac{\Delta}{\omega_c})^{\alpha/(1-\alpha)}$ for $T \lesssim \Delta_r$. The quality factor of the oscillations agrees with predictions from the Non Interacting Blip Approximation (NIBA)~\cite{RevModPhys.59.1}, field theory~\cite{PhysRevLett.80.4370} and from the time-dependent numerical renormalization group (TD-NRG)~\cite{roosen_hofstetter_lehur_unpublished}.

For intermediate values of alpha we are able to access the asymptotic long-time behavior of $P(t)$, where $|P(t)| \ll 1$, within our numerical approach. At $T=0$, we find that the system exhibits exponentially damped coherent oscillations as predicted in Ref.~\cite{PhysRevLett.80.4370} and in agreement with recent TD-NRG calculations~\cite{roosen_hofstetter_lehur_unpublished}. Note that the statistical error of our method scales like $N^{-1/2}$, where $N$ is the number of different realizations of the noise. We thus cannot resolve the existence of a small incoherent part at weak-coupling $\alpha \ln \frac{\omega_c}{\Delta} \ll 1$, which was predicted in Ref.~\onlinecite{PhysRevB.71.035318} using a rigorous Born approximation.
\begin{figure*}[t!]
  \centering
\includegraphics[width=.7\linewidth]{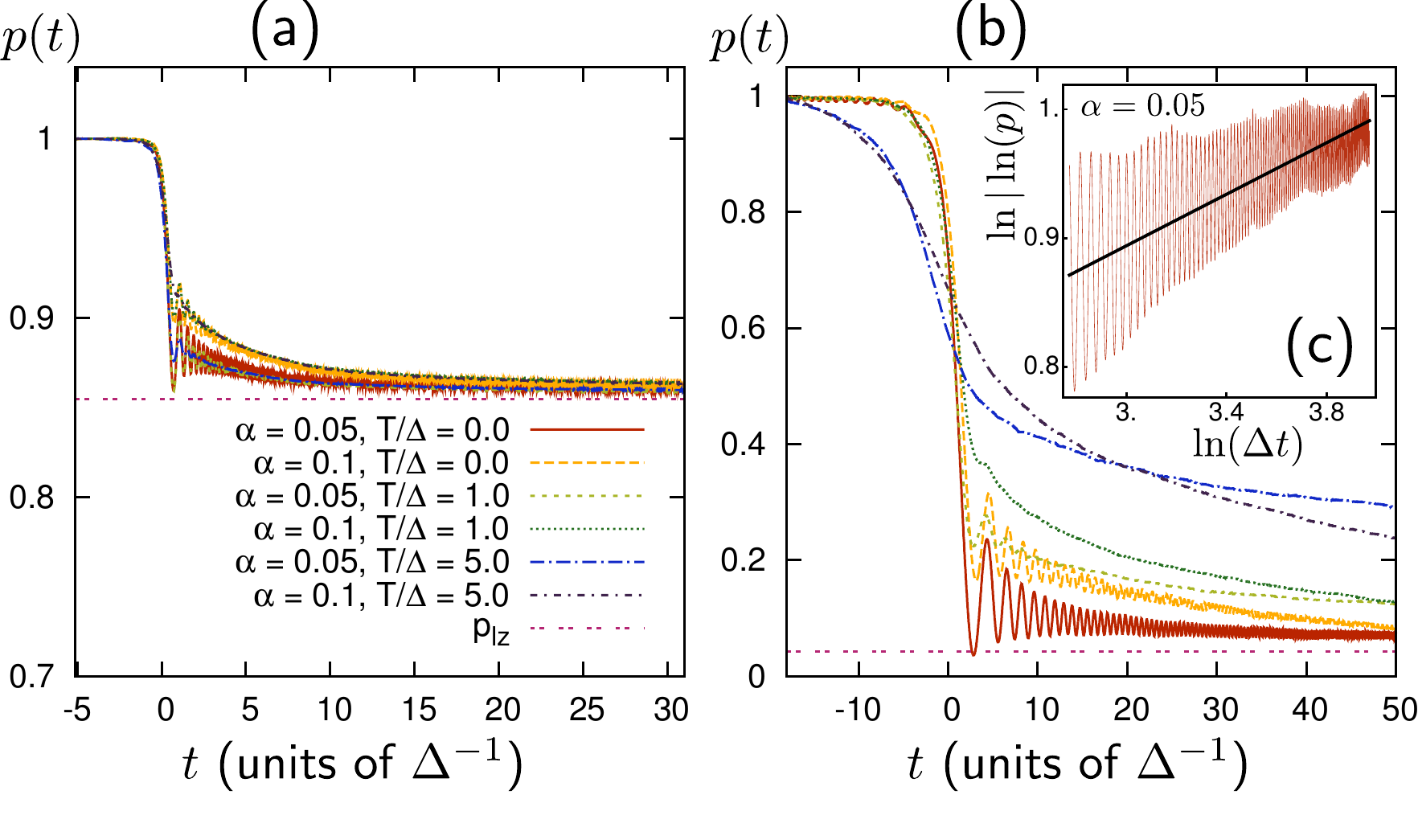}
  \caption{(Color online) (a) $p(t)$ for a fast sweep with $v/\Delta^2=10$, $\omega_c/\Delta = 200$, $\alpha = \{0.05, 0.1\}$ and $T/\Delta=\{0,1,5\}$ (here, $\hbar = k_B =1$). We choose $m_{\max}=4000$, $N=4 \cdot 10^6$. (b) Slow sweep with $v/\Delta^2=0.5$. Other parameters as in (a). (c) Fit of universal decay of $p[\epsilon(t)]$ using Eq.~\eqref{eq:14} with $\alpha=0.05$ and single fit parameter $C=0.59$.}
  \label{fig:3}
\end{figure*}

For increasing temperature, the coherence of oscillations gets lost more rapidly, and finally for $T \gg \Delta_r$ we observe incoherent decay $P(t) = \exp[- t/\tau]$ with rate $\tau^{-1} = \frac{\sqrt{\pi}\Gamma(\alpha)}{2 \Gamma(\alpha + \frac12)} \frac{\Delta_r^2}{T} \bigl[ \frac{\pi T}{\Delta_r} \bigr]^{2 \alpha}$~\cite{RevModPhys.59.1}. Note that our method gives reliable results over the full range of temperatures. 
\subsection{Dissipative Landau-Zener transition}
\label{sec:land-zener-trans}
Next, we turn to the case of a Landau-Zener sweep of the detuning $\epsilon(t) = v t$ with $v>0$. We examine the survival probability $p(t)$ that the spin remains in its initial state if swept across the resonance. Neglecting the bath, this problem can be solved exactly~\cite{landau_lz,zener_lz,stueckelberg_lz,majorana_lz} and one finds that $p(t)$ converges toward the celebrated Landau-Zener formula $p_{lz} = \exp [ - \pi \Delta^2/2v]$ for $t \gg \Delta/v$.

A fundamental question is thus how this result is modified in the presence of dissipation. Surprisingly, at zero temperature the bath does not affect the final transition probability $p_{lz}$ in the limit $t\rightarrow +\infty$, if the spin couples longitudinally to the reservoir via its $\sigma^z$ component~\cite{wubs:200404}. This limit, however, corresponds to very large times $t\gg \omega_c/v$ where the separation of the spin energies is larger than the bosonic bandwidth. In contrast, we explore the so-called {\it scaling} regime, where one first takes the limit $\omega_c \rightarrow \infty$, holding $\Delta_r t = y$ fixed, and only then considers $y\rightarrow \infty$. This limit is important because it allows the spin-boson model to exhibit universal behavior~\cite{RevModPhys.59.1,weissdissipation}. For large but finite $\omega_c$ the scaling regime corresponds to an \emph{intermediate time regime} where the spin energy separation $\epsilon$ is smaller than $\omega_c$ but possibly much larger than $\Delta$: $\Delta\ll v t\ll \omega_c$. Phonon assisted spin transitions therefore still occur even though $\epsilon \gg \Delta$, and the probability $p(t)$ converges toward its final value $p_{lz}$ only for times of the order $t\sim \omega_c/v$. Note that this is in stark contrast to the non-dissipative (perfectly isolated) case where this convergence happens much faster for $t \sim \Delta/v$. 

In the context of Landau-Zener transitions, the bath preparation affects the long-time result of $p(t)$~\cite{saito:214308}. Thus, it is important to consider the contribution of the initial sojourn exactly, 
as it accounts for the fact that the bath starts out from a shifted equilibrium state. 
It is given by the $k=0,1$ terms in $\mathcal{Q}_1$ (Eq.~\eqref{eq:4}). We can incorporate this term by adding it to the height function
\begin{align}
  h(\tau, \tau_1) & = \frac{v t_{\max}^2}{2} (\tau^2 - 2 \tau_c \tau) + \sum_{m=1}^{m_{\max}} s_m \sqrt{ - \alpha G_m} \Psi_m(\tau) \notag \\ & - 2 \alpha \tan^{-1}[ \omega_c t_{\max} ( \tau - \tau_1)]\,. \label{eq:13}
\end{align}
Here, $t_{\max}$ determines the time interval length of our simulation, and $[0,\tau_c]$ $([\tau_c, 1])$ correspond to times before (after) the level crossing.
The fact that the height function now contains $\tau_1$ forces us to explicitly perform the $\tau_1$-integration in Eq.~\eqref{eq:8}. We thus randomly pick a uniformly distributed $\tau_1 \in [0,1]$, which determines $h(\tau, \tau_1)$ as well as the initial state $\ket{\Phi_{\tau_{1}}} = - i (0, e^{i h(\tau_1, \tau_1)}, - e^{- i h(\tau_1,\tau_1)}, 0)^T$. We then propagate this initial state in the interval $[\tau_1, 1]$ according to Eq.~(\ref{eq:10}) and calculate the survival probability as $p(\tau) = 1 + \av{\Phi_1(\tau)}$, where the average is over $N$ choices of $\tau_1$ and random variables $\{s_m\}$. Here we set $\ket{\Phi(\tau < \tau_1)}=0$ in an individual run since $\av{\Phi_1(\tau)}$ only accounts for the contribution of paths with at least one spin jump. In Eq.~(\ref{eq:10}), the evolution is not unitary.


In Fig.~\ref{fig:3}~\!(a), we check that $p(t)$ converges toward $p_{lz}$ at long times $t \gg \omega_c/v$ for $T=0$. For the large sweeping speed $v/\Delta^2 = 10$ in Fig.~\ref{fig:3}~\!(a), we find that this also holds for $T/\Delta=\{1, 5\}$, since thermal effects only occur during the short period where $|v t| \lesssim T$~\cite{nalbach:220401}. In Fig.~\ref{fig:3}~\!(b), we show results for slow sweep velocity $v/\Delta^2 = 0.5$, and observe clearly that the size of the jump in $p(t)$ at the crossing reduces with enhancing dissipation and temperature. Following the jump, we observe a decay of $p(t)$ in the intermediate time regime up to $t \sim \omega_c/v$ due to bath mediated spin transitions. The final probability increases with temperature due to thermalization.  

We now derive an analytical formula describing the universal decay in the scaling regime at $T=0$, which holds for slow sweeping speeds only. For large static detuning $\epsilon \gg \Delta_r$ (but still $\epsilon \ll \omega_c$) the NIBA can be justified~\cite{RevModPhys.59.1} and predicts an overdamped exponential relaxation with a decay rate $\Gamma= \frac{\pi \Delta_r}{2 \Gamma(2 \alpha)} (\epsilon/\Delta_r)^{2 \alpha - 1}$. Inserting $\epsilon(t)=v t$ and integrating $dp/dt = - \Gamma p(t)$ for $\alpha< 1/2$ yields
\begin{equation}
\label{eq:14}
p[\epsilon(t)] = C \exp \Bigl[ \frac{ - \pi \Delta_r^2}{ 4 \alpha \Gamma(2\alpha) v} \Bigl( \frac{\epsilon}{\Delta_r} \Bigr)^{2 \alpha} \Bigr]\,.
\end{equation}
If we except the integration constant $C$, this formula contains only scaling variables, which shows that the decay is universal. It reduces to $p_{lz}$ in the limit $\alpha \rightarrow 0$ (with $C =1$) and breaks down for times of the order $t \sim \omega_c/v$, where it becomes a function of the bare $\Delta$ again: $p[\epsilon=\omega_c] = C \exp[- \frac{\pi \Delta^2}{4 \alpha \Gamma(2\alpha) v}]$.  In Fig.~\ref{fig:3}~\!(c), we show that the decay is indeed described by this formula for $T=0$, $v/\Delta^2 = 0.5$ and $\alpha = 0.05$. Note that this decay does not occur at $\alpha=1/2$~\cite{PhysRevB.79.115137}. 
We like to emphasize that our numerical method gives reliable results over the whole range of sweep velocities and temperatures. 

\subsection{Realization with cold-atom quantum dot setup}
\label{sec:real-with-cold}
The intermediate (scaling) time regime $\Delta \ll v t \ll \omega_c$ might be accessed using the cold-atom geometry of Refs. \cite{recati:040404,orth:051601,PhysRevLett.104.200402}. It comprises a bosonic mixture of atoms in two hyperfine ground states $a$ and $b$, subject to state-selective traps. One species forms a one-dimensional Bose-Einstein Condensate (BEC), representing the Ohmic reservoir, and the other species is trapped in a tight harmonic potential, operated in the collisional blockade limit, representing the ``spin''. Coupling the different species by Raman lasers, the system is described by Eq.~(\ref{eq:1}) with $\Delta$ and $\epsilon$ being proportional to the laser intensity and frequency, respectively. Using the parameters of Ref.~\cite{ToshiyaKinoshita08202004}, we estimate
$\alpha = \frac{1}{4 K} (-1 + g_{ab}/g_{aa})^2 \approx 0.06$; $K\sim \sqrt{\rho_a/g_{aa}}$ is the Luttinger parameter of the BEC, $g_{\alpha\beta} = 2 \hbar \omega_\perp a_{\alpha \beta}$ are the scattering amplitudes containing the transverse trapping frequency $\omega_\perp= 2 \pi \times 67 \text{kHz}$ and the scattering length $a_{aa}=5.2 \text{nm}$. The value of $a_{ab}$ must be tuned such that $g_{ab} \ll g_{aa}$ using optical Feshbach resonances~\cite{RevModPhys.82.1225}. Choosing $\Delta \approx 100 \text{Hz}$ and $v\approx 1 \text{kHz}/\text{s}$, the intermediate time (scaling) regime occurs between $0.1 \text{s} < t < 10 \text{s}$.

\section{Conclusions}
\label{sec:conclusions}
To summarize, we have developed a stochastic Schr\"odinger method to investigate the dissipative Landau-Zener problem in the scaling limit $\Delta/\omega_c \ll 1$ at finite temperature. Assuming $\alpha<1/2$, we have shed light on an experimentally relevant intermediate time-regime where $p(t)$ shows universal decay due to bath mediated spin transitions. Our results are relevant in quantum information, where fast quantum processes are more useful. Our method can also be extended to other many-body environments. 

\section{Acknowledgments}
\label{sec:acknowledgments}
We acknowledge discussions with L. Glazman, W. Hofstetter, and D. Roosen. This work was supported by the U.S. Department of Energy, Office of Basic Energy
Sciences, Division of Materials Sciences and Engineering under Award
DE-FG02-08ER46541 and by the Yale Center
for Quantum Information Physics through the grant NSF DMR-0653377.
%

\end{document}